# Propagation-invariant high-dimensional orbital angular momentum states

**Li-Wei Mao[1], Dong-Sheng Ding[1,2], Carmelo Rosales-Guzmán[1,3], and Zhi-Han Zhu[1]**

[1] Wang Da-Heng Center, Heilongjiang Key Laboratory of Quantum Control, Harbin University of Science and Technology, Harbin 150080, China
[2] CAS Key Laboratory of Quantum Information, University of Science and Technology of China, Hefei, 230026, China
[3] Centro de Investigaciones en Óptica, A.C., Loma del Bosque 115, Colonia Lomas del campestre, 37150 León, Gto., Mexico

E-mail: carmelorosalesg@cio.mx; zhuzhihan@hrbust.edu.cn

**Abstract**

Photonic states encoded in spatial modes of paraxial light fields provide a promising platform for high-dimensional quantum information protocols and related studies, where several pioneering theoretical and experimental demonstrations have paved the path for future technologies. Crucially, critical issues encountered in free-space propagation still represent a major challenge. This is the case of asynchronous diffraction between spatial modes with different modal orders, which experience variations in their transverse structure upon free-space propagation. Here we address this issue by proposing an encoding method based on the use of Laguerre-Gaussian (LG) modes of the same modal order $N$ to define a $N + 1$ dimensional space. Noteworthy, such modes endowed with orbital angular momentum (OAM) experience the same propagation aberrations featuring an identical Gouy phase and wavefront curvature. We demonstrate our proposal experimentally by using time-correlated-single-photon imaging combined with a digital propagation technique. Importantly, our technique allows to eliminate, without the use of imaging systems, all issues related to asynchronous diffraction, providing an accessible way to generate propagation-invariant OAM qudits for quantum optical protocols.

## 1. Introduction

With the development of quantum science over the past century, several advances have revolved around the use of quantum states and their dynamic behavior. Among the various possible quantum states, those constructed from photons and their associated states have played significant roles, from the dawn of quantum science to the recent quantum-information revolution [1,2]. Driven by a continuous advance in photonic techniques, the generation and control of quantum states encoded with photonic degrees of freedom (DOFs) are now readily available, many of which have been converted to an integrated form [3]. On this basis and from an application perspective, scientists are trying to build a global quantum network [4,5]. Simultaneously, the advantages of photonic systems in quantum computer technologies, for the development of specific tasks, have been demonstrated [6].

Noteworthy, quantum states constructed in a high-dimensional Hilbert space (qudits) can provide a larger encoding space and have advantages beyond the 2D qubit in quantum information processing [7-10]. Thus, exploring novel ways to generate and control photonic states with higher dimensionalities and larger photon numbers has become a topic of late.

High-dimension photonic states have been previously considered in the path and frequency-time DOFs [11,12] and more recently have focused on the transverse structure of photon (or laser) beams, including the spatial modes of paraxial light fields [13,14]. Fueled by the rapid progress in the field of structured light over the past three decades, there have been developed various techniques and toolkits to shape and characterize the spatial amplitude, phase, and polarization of light fields [15-17]. Importantly, by exploiting unbounded spatial modes, one can encode and control high-dimension states within a paraxial beam. Within this context several promising quantum experiments have been taken from a constrained lab environment to a more realistic outdoor demonstration. [18-27]. In relevant studies, it has become traditional the use of spatial modes that are eigen solutions of the orbital angular momentum (OAM), such as the Laguerre–Gaussian (LG) and Hyper-Geometric-Gaussian (HyGG) modes [13], and the associated vector modes that feature spin-orbit coupling (SOC) [28-31]. For the interested reader, there are several comprehensive reviews [32-37]. In this work we address a common issue related to the dependency of structured-photon control with propagation distance, which originates from the asynchronous diffraction between spatial modes with different modal orders.

Measurement and intermodal interference of spatial modes are crucial ways to implement quantum control of high-dimensional states encoded on structured photons. As the diffraction upon propagation experienced by higher-order optical modes depends on their modal number, each may diffract at different rates, therefore, quantum control is sensitive to wavefront of structure photons, i.e., measurement and interference of spatial modes becomes propagation dependent. In consequence, imaging systems have been used in most demonstrations of free-space links to eliminate the influence of diffraction, which limits the advantages offered by structured photons towards large-scale and outdoor applications. For the OAM DOF, only two-Dimensional (2D) states ($d = 2$) encoded using conjugate OAM modes are in principle propagation invariant states. Although one can use a hybrid (direct product) space of the OAM and polarization to break this limitation, i.e., SOC space, the maximum achievable dimensionality is $d = 4$ [26,27]. This drawback eliminates the advantages of OAM states with $d > 2$ in many tasks. To illustrate this, in the following section we provide a concrete example.

## 2. Methods & Results

In what follows we will restrict ourselves to the set of LG modes, whose spatial wavefunction in cylindrical coordinates $\{r, \varphi, z\}$ is given by [28],

$$LG_p^\ell(r,\varphi,z) = \sqrt{\frac{2p!}{\pi(p+|\ell|)!}} \frac{1}{w_z} \left(\frac{\sqrt{2}r}{w_z}\right)^{|\ell|} \exp\left(\frac{-r^2}{w_z^2}\right) \times L_p^{|\ell|}\left(\frac{2r^2}{w_z^2}\right) \exp\left[-i\left(kz + \frac{kr^2}{2R_z} + \ell\varphi - \phi_g\right)\right], \quad (1)$$

where $\ell$ ($p$) denote the azimuthal (radial) index and $L_p^{|\ell|}(\cdot)$ represent the associated Laguerre polynomials. Compared with propagation-variant HyGG modes, the propagation of LG modes is elegant with self-similar beam profiles whose diameter increases as $w_z = w_0\sqrt{1+(z/z_R)^2}$. In addition, there are two parameters that govern the diffraction behavior of LG modes upon propagation, namely, the Gouy phase and the radius of curvature, $\phi_g = (2p + |\ell| + 1)\arctan(z/z_R)$ and, $R_z = z^2 + z_R^2/z$, respectively, where $z_R = kw_0^2/2$ is known as the Rayleigh length.

To illustrate the effect of asynchronous diffraction, we consider 3D states constructed from a subset of LG modes and their superposition given by azimuthal index $\ell = 0, 1, 2$ and radial number $p = 0$, which for convenience we denote with the ket $|\ell\rangle$. In the Hilbert space, there are $d + 1 = 4$ mutually unbiased bases (MUBs) that can be interconverted using 3D Hadamard transformations, which we denoted as MUB-n (n = I, II, III, and IV),

I: $\{|0\rangle, |1\rangle, |2\rangle\}$

II: $\left\{\frac{|0\rangle+|1\rangle+|2\rangle}{\sqrt{3}}, \frac{|0\rangle+\omega|1\rangle+\omega^2|2\rangle}{\sqrt{3}}, \frac{|0\rangle+\omega^2|1\rangle+\omega|2\rangle}{\sqrt{3}}\right\}$

III: $\left\{\frac{|0\rangle+\omega|1\rangle+\omega|2\rangle}{\sqrt{3}}, \frac{|0\rangle+\omega^2|1\rangle+|2\rangle}{\sqrt{3}}, \frac{|0\rangle+|1\rangle+\omega^2|2\rangle}{\sqrt{3}}\right\}$ (2)

IV: $\left\{\frac{|0\rangle+\omega^2|1\rangle+\omega^2|2\rangle}{\sqrt{3}}, \frac{|0\rangle+|1\rangle+\omega|2\rangle}{\sqrt{3}}, \frac{|0\rangle+\omega|1\rangle+|2\rangle}{\sqrt{3}}\right\}$

where $\omega = \exp(i2\pi/3)$ for $d = 3$. Upon propagation in free space, the spatial structure of all the modes in the basis MUB-I remains invariant, except for an overall enlargement. Nonetheless, each of the modes in the three remaining basis changes its spatial structure since each mode has a different Gouy phase $\phi_g$, for example the first state in MUB-II becomes $1/\sqrt{3}\left(|0\rangle + e^{i\arctan(z/z_R)}|1\rangle + e^{i2\arctan(z/z_R)}|2\rangle\right)$, which clearly indicates a z-dependence on the intermodal phases [38,39]. Thus, the orthogonality relation in this space, except for MUB-I, is only valid for $z = 0$. One way to maintain the high dimensionality is through the use of imaging system. Additionally, the coherent manipulation of spatial modes is wavefront sensitive, this is the case of mode measurements and intermodal interference. Therefore, it is crucial to ensure all the modes under a coherent operation have an identical wavefront curvature. However, the wavefront curvature accumulated from diffraction is also asynchronous for different mode orders $N = 2p + |\ell|$.

To demonstrate the influence of the above two unwanted effects, we first inspect the orthogonality relation of the modes in Eq. (2) as function of the propagation distance. Figure 1(a)-(c) shows the theoretical projections between the states in all MUBs calculated at the planes $z_0$, $z_R$, and $z_f$. Here, the projection is defined as the inner product between the modes in the MUB basis. For example, the projection of mode $|1\rangle$ in the modes of the MUB-I is performed as

$$\eta = \iint LG_0^1(r,\varphi,z)LG_0^{-\ell}(r',\varphi,z_0)rdrd\varphi, \quad (3)$$

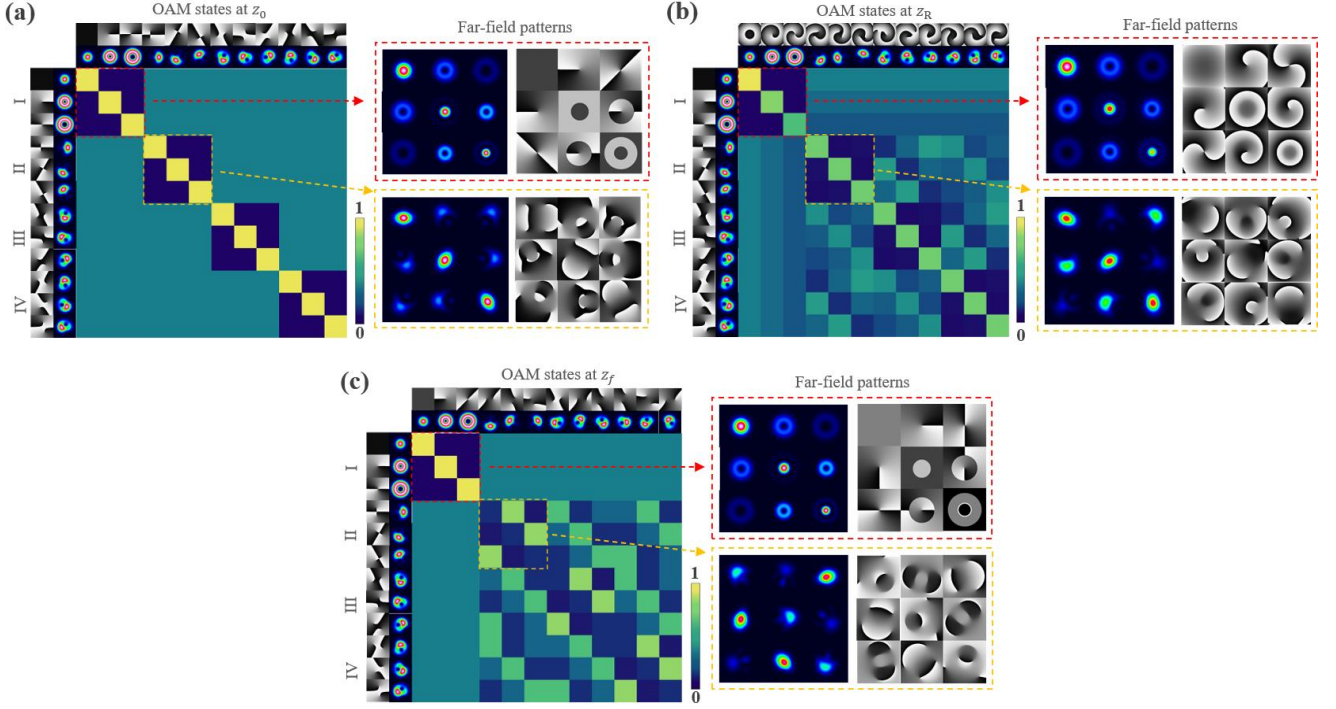

Figure 1. Simulated correlation matrices between MUBs versus diffraction distance, where patterns in dotted-line boxes are far-field complex amplitudes of measured states marked in the matrices.

where $LG_0^{-\ell}(r',\varphi,z_0)$ represents the complex-conjugated spatial complex amplitudes of the modes in MUB-I with a rescaled radial coordinate $r' = r/\sqrt{1+(z/z_R)^2}$ (to match the beam size of the measured mode). This projective integral can be realized experimentally using complex-amplitude modulation holograms [19], and the measured result is proportional to the on-axis intensity of the far-field pattern [15]. Figure 1(a) indicates that (i) the perfect orthogonal relation can only be achieved at the plane $z_0$, where $\phi_g = 0$ and $R_z \to \infty$ for all modes; (ii) at the Fourier plane $z_f$, as shown in Fig. 1(c), although $R_z \to \infty$ for all modes, the mode-dependent $\phi_g$ disturbs the intermodal phase relations between the modes of the last three MUBs; and (iii) the worst case occurs the modes propagate to the far-field (such as at the plane $z_R$), as shown in Fig. 1(b), in which both $R_z$ and $\phi_g$ are mode-dependent for all modes. Even for the projections within MUB-I, the spherical wavefront exists in the far-field patterns, which breaks the uniformity in projective efficiency.

We now further inspect the intermodal interference $1/\sqrt{3}(|0\rangle + e^{i\theta_1}|1\rangle + e^{i\theta_2}|2\rangle)$ obtained at the three $z$ planes, which mimics a three-path interferometer of different spatial modes. This can be implemented experimentally by simultaneously measuring the three modes, which leads to a 2D probability distribution of $\mu(\theta_1, \theta_2) = 1/\sqrt{3}(1 + e^{i\theta_1} + e^{i\theta_2})$.

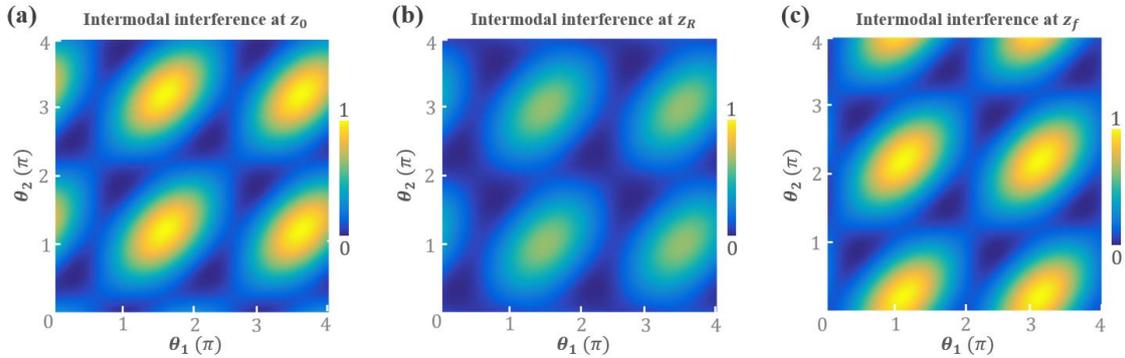

Figure 2. Simulated probability distributions of intermodal interference observed at the planes (a) $z_0$, (b) $z_R$, and (c) $z_f$. The calculation was based on a 3D state $1/\sqrt{3}(LG_0^0 + e^{i\theta_1}LG_0^{+1} + e^{i\theta_2}LG_0^{+2})$ with $z = z_0$, $z_R$, and $\infty$, respectively.

Crucially, while in a real interferometer, the multi-mode phase relationship may be unstable, in this interferometer-like superposition of OAM modes, the phase relation can be highly stable. Nonetheless, as shown in Fig. 2, this stability can only be achieved with the help of imaging systems. To be more precise, (i) upon propagation, *i.e.*, for planes $z>0$, the probability distribution shifts globally in both directions ($\theta_1$ and $\theta_2$) due to the dependency of the Gouy phase ($\phi_g$) with the modal number *N*, which in principle can be corrected before the experiments; (ii) in addition, the parameter $R_z$, which is different for each OAM mode, also contributes to changes in the shape of the probability distribution, its correction is extremely difficult. Crucially, the effects produced by the mode-dependent radius of curvature ($R_z$) represent a drawback in many optical experiments, such as the interference between photons. More details regarding the influence of $R_z$ on the behaviors of spatial modes in coherent operations are beyond the scope of this paper and will be discussed in a future paper.

*The influence of an asynchronous mode-dependent behavior of both, $\phi_g$ and $R_z$ is difficult to overcome, nonetheless it can be smartly evaded.* Following the above discussion, a straightforward solution is to give all the involved spatial modes an identical mode order *N*. Of note, the azimuthal indices ($\ell$) of LG modes are just eigen values of the paraxial OAM. For a given mode order *N*, there are a total of $N+1$ different eigenvalues, i.e., $\ell = -N, -N+2, ..., N-2, N$, that correspond to $N+1$ orthogonal LG modes [40-42]. Thus, LG modes of the same mode order *N* can be used to encode high-dimension states of dimension $d = N+1$. To achieve this, we need to exploit both the azimuthal and radial indices of the LG modes. The research community has made great efforts in relevant studies recently in this regard [24, 43-47]. The use of the direct product space of the azimuthal and radial indices can drastically boost the dimensionality, which is much higher than using the SOC space. Notice that the values of $\ell$ and *p* have to be carefully chosen according to the relation $N = 2p + |\ell|$. Hence, here the dimensionality is decided only by the azimuthal index, while the radial index is chosen to achieve the same modal number, and therefore to ensure the modes propagate in an identical way [37,38]. In other words, the qudits is encoded using both transverse indices that completely define the transverse structure of LG modes [48,49].

The remaining task is to define the full MUB sets using all the LG modes with the same modal number. Importantly, although the generalized Hermite–Laguerre–Gaussian modes can form a complete modal sphere with an SU(2) structure, the MUB relationship between the HG and LG modes is only established for $N = 1$. Thus, to construct the full MUB sets in a high dimensional space of $d > 2$ ($N > 1$), we require to use high-dimensional Hadamard transformation. For example, for $d = 4$ ($N = 3$) this can be obtained through the matrix representation,

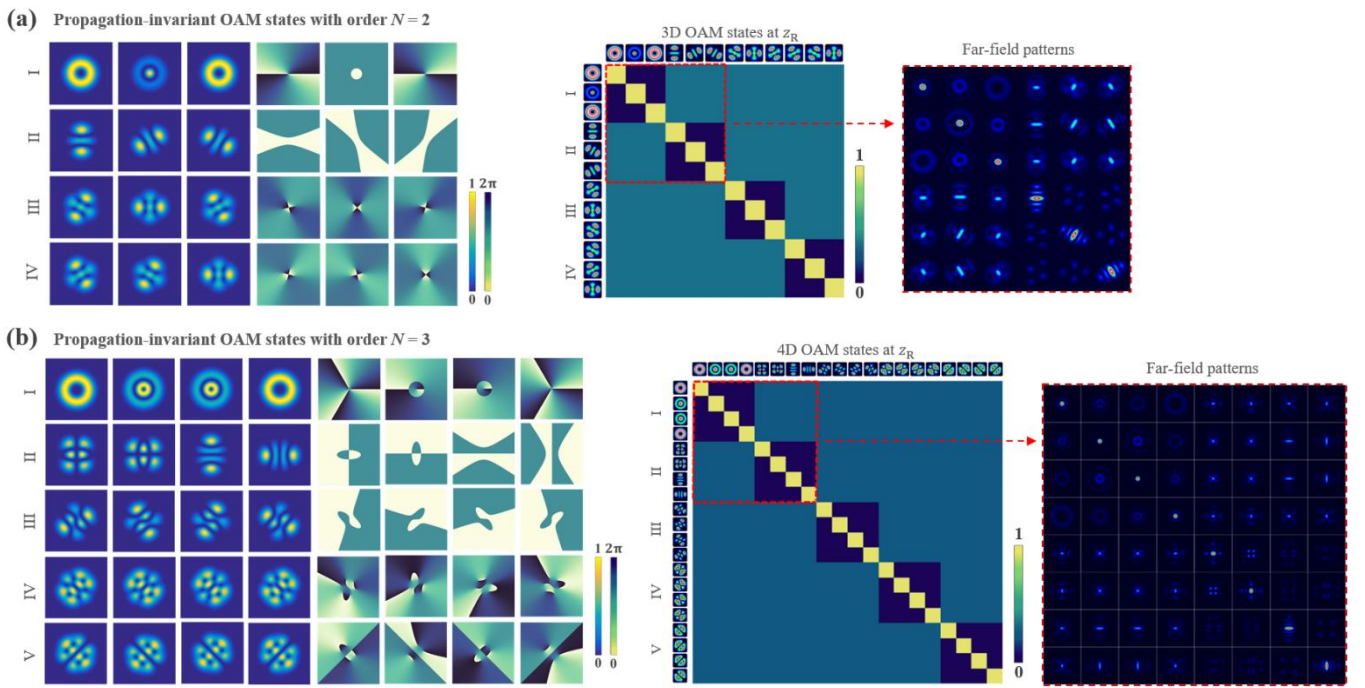

Figure 3. Simulated complex amplitudes of propagation-invariant OAM states with (a) $N = 2$ and (b) $N = 3$, as well as correlation matrices between all MUBs obtained at the $z_R$ plane.

$$\hat{H}_1 = \frac{1}{2}\begin{bmatrix} 1 & 1 & 1 & 1 \\ 1 & -1 & 1 & -1 \\ 1 & 1 & -1 & -1 \\ 1 & -1 & -1 & 1 \end{bmatrix}; \quad \hat{H}_2 = \frac{1}{2}\begin{bmatrix} 1 & 1 & 1 & 1 \\ i & -i & i & -i \\ i & i & -i & -i \\ -1 & 1 & 1 & -1 \end{bmatrix};$$

$$\hat{H}_3 = \frac{1}{2}\begin{bmatrix} 1 & 1 & 1 & 1 \\ i & -i & i & -i \\ 1 & 1 & -1 & -1 \\ -i & i & i & -i \end{bmatrix}; \quad \hat{H}_4 = \frac{1}{2}\begin{bmatrix} 1 & 1 & 1 & 1 \\ 1 & -1 & 1 & -1 \\ i & i & -i & -i \\ -i & i & i & -i \end{bmatrix}. \quad (4)$$

Figure 3 shows the simulated complex amplitudes for all states at the $z_0$ plane, and corresponding correlation matrices between all MUBs obtained at the plane $z_R$ for $N = 2$ (Fig. 3a) and $N = 3$ (Fig. 3b).

We now demonstrate experimentally the proposed method using the setup shown schematically in Fig. 4. The experiment utilizes a type-II spontaneous parametric down-conversion (SPDC) as the heralded photon source. The SPDC occurred in a 10 mm PPKTP crystal pumped with a 397.5 nm laser, which was the second-harmonic wave of a 795 nm laser obtained in another crystal (10 mm type-0 PPKTP). The SPDC photon pairs were sorted using a polarizing beam splitter (PBS), one arm was directly converted into the TTL signal to trigger a single-photon sensitive camera (EMICCD), and the other was relayed through a 30 m single-mode fiber (SMF) and converted into a TEM$_{00}$ mode, as a signal for the time-correlated-single-photon imaging. In the free-space link, a group of wave plates combined with a PBS converted the signal photons into horizontal polarization. Then, the two spatial light modulations (SLM) combined with two Fourier lenses constructed a digital holography system that we used for generation and characterization spatial modes. Specifically, SLM-1 first converted the signal photons into the target modes, such as those patterns shown in Fig. 3, through complex-amplitude modulation [16, 50]. In addition, a digital-propagation phase mask was multiplexed on the hologram of complex-amplitude modulation to control the equivalent propagation distance of generated modes arriving at the surface of SLM-2, i.e., the Fourier plane of L1. Then, SLM-2 converted one (or several) mode(s) into a Gaussian mode through complex-amplitude modulation and a single-photon sensitive camera (EMICCD) recorded the converted patterns at the Fourier plane of L2 [51].

The experiments set an identical propagation distance $z = z_R$ for all states to be measured. Compared with the results in Fig. 1(b), we first examined the MUB relationships for 3D space as defined by $\{LG_0^{-2}, LG_1^0, LG_0^{+2}\}$. Figure 5(a) shows the measured projections between the azimuthal-radial coupled states over all MUBs. The results agree well with the MUB relationships as given by $|\langle i_m | i_m \rangle|^2 = 1$, $|\langle i_m | i_n \rangle|^2 = 0$, and $|\langle i | j \rangle|^2 = 1/3$. This indicates that the problem induced by asynchronous intermodal diffraction no longer exist. For clarity, we measured the 'three-path interference' of $1/\sqrt{3}(|-2\rangle + e^{i\theta_1}|0\rangle + e^{i\theta_2}|+2\rangle)$, as shown in Fig. 5(b), and the observed 2D probability distribution perfectly matched the theoretical distribution of an ideal (propagation independent) 3D state. Similarly, we further examined the encoding method in a 4D space defined by $\{LG_0^{-3}, LG_1^{-1}, LG_1^{+1}, LG_0^{+3}\}$, where the states over all MUBs were constructed using Eq. (4). The corresponding results shown in Figs. 5(c) verify the advantage of the proposed encoding method, which, in principle, can be further extended into vectorially spatial modes.

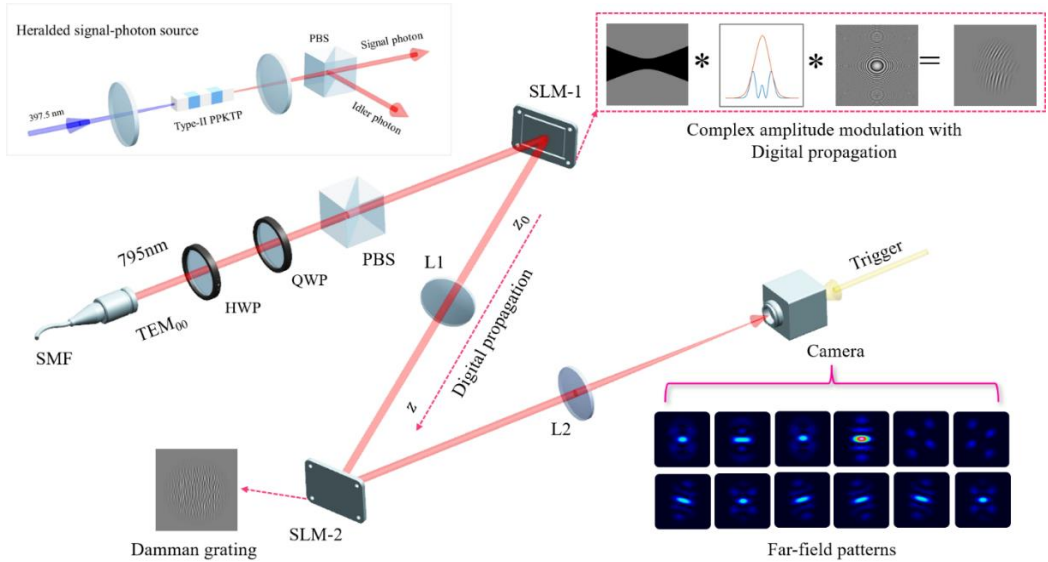

Figure 4. Diagram of the experimental setup, where the key components include the single-mode fiber (SMF), lens (L1/2), polarizing beam splitter (PBS), half- / quarter-wave plate (HWP/QWP), spatial light modulator (SLM), and Camera.

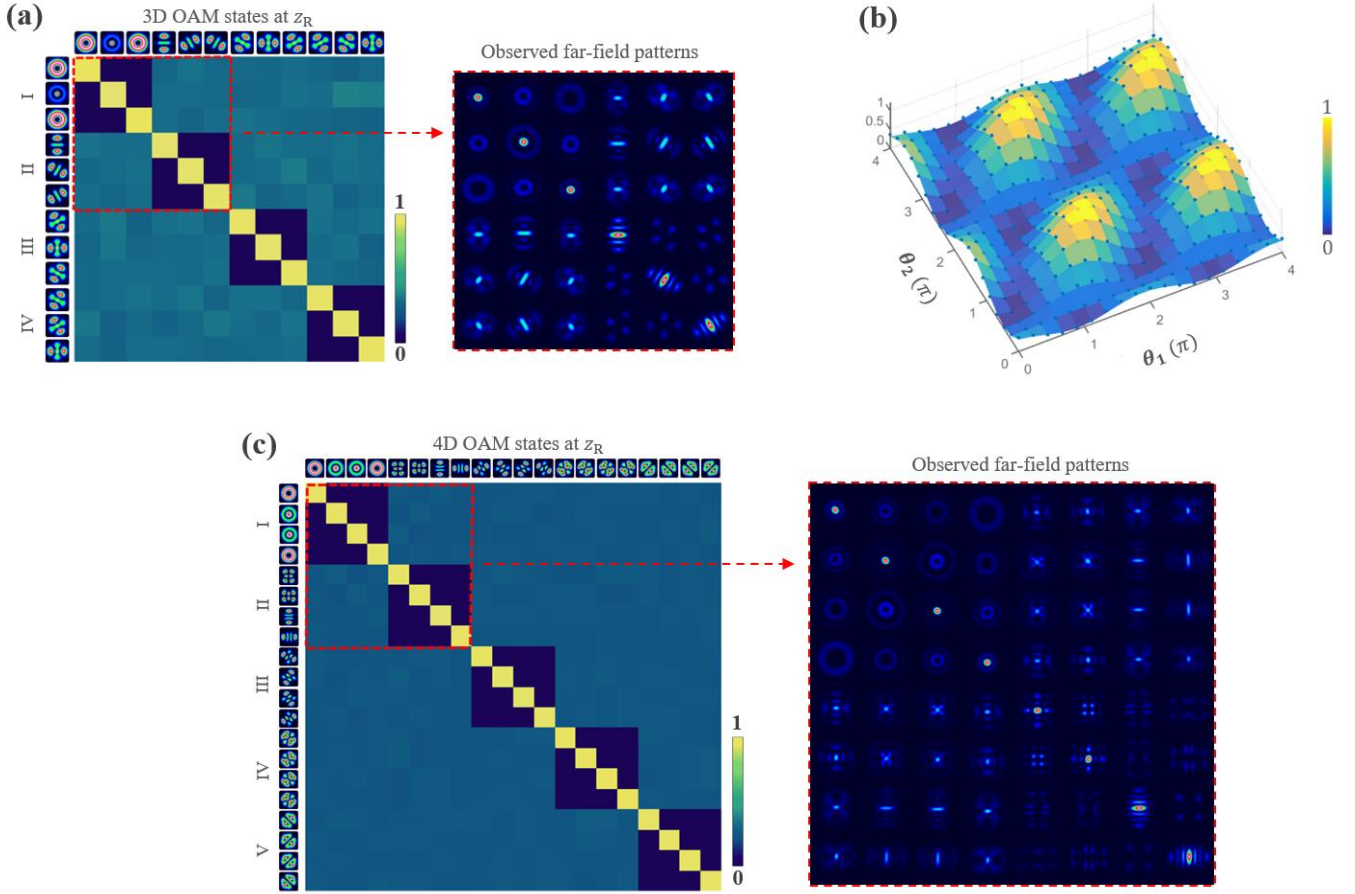

Figure 5. Experimentally measured correlation matrices of high-dimensional states with (a) $N = 2$ and (c) $N = 3$ obtained at the $z_R$ plane, as well as (b) the 'three-path' intermodal interference.

A significant challenge to date is the lack of compact geometric-phase elements that enable spin-dependent complex-amplitude modulations [52,53]. Moreover, the measurement we adopted in the demonstration was filtering-based projective measurement, which means the loss rate of photons increases with the dimensionality. To date, the community of structured light has made significant efforts in sorter-based (lossless) measurement for spatial modes, including both azimuthal and radial modal sorters [54-57]. For the proposed propagation-invariant OAM states, the sorter-based measurement requires a radial-mode-independent OAM sorter, which seems already available but still needs to be further examined.

## 3. Conclusion

This paper focuses on a critical issue encountered when using photonic spatial modes to realize experimentally high-dimensional quantum information. A detailed analysis shows that asynchronous diffraction existing between spatial modes of different orders leads to variations in the wavefront curvature and intramodal phase of the modes upon propagation. This prevents high-dimensional states encoded with complex structured light from moving toward larger scales and outfield applications without the help of an imaging system. To address this issue, we proposed an encoding method using LG modes of order $N = 2p + |\ell|$ to define a $d = N + 1$ dimensional space, successfully removing the problems associated to asynchronous diffraction. That is, the information is encoded in the azimuthal-radial coupled modes with an identical order $N$. All the MUBs in this space can be easily found using a high-dimensional Hadamard transformation. A triggered single-photon imaging device combined with a digital-propagation technique experimentally verified the proposed method. Using this method, one can easily generate arbitrarily dimensional OAM qudits with propagation-invariant structure, which reduce the complexity of free-space quantum optical experiments.

**Acknowledgements**

This work was supported by the National Natural Science Foundation of China (Grant Nos. 62075050, 11934013, and 61975047) and the High-Level Talents Project of Heilongjiang Province (Grant No. 2020GSP12).